# Real-World Data Analysis of Implantable Cardioverter Defibrillator (ICD) in Patients with Hypertrophic Cardiomyopathy (HCM)

Sungrim Moon, PhD [a], Andrew Wen, MS [a], Christopher G. Scott, MS [a], Michael J. Ackerman, MD, PhD [b], Jeffrey B. Geske, MD [b], Peter A. Noseworthy, MD [b], Steve R Ommen, MD [b], Jane L Shellum [c], Hongfang Liu, PhD [a], Rick A. Nishimura MD [b]

Department of Health Sciences Research [a], Department of Cardiovascular Diseases Medicine [b], Robert and Patricia Kern Center for Science of Health Care Delivery [c], Mayo Clinic, Rochester, MN, USA

## Abstract

**Background**: One of the common causes of sudden cardiac death (SCD) in young people is hypertrophic cardiomyopathy (HCM) and the primary prevention of SCD is with an implantable cardioverter defibrillators (ICD). Concerning the incidence of appropriate ICD therapy and the complications associated with ICD implantation and discharge, patients with implanted ICDs are closely monitored and interrogation reports are generated from clinical consultations.

**Methods:** In this study, we compared the performance of structured device data and unstructured interrogation reports for extracting information of ICD therapy and heart rhythm. We sampled 687 reports with a gold standard generated through manual chart review. A rule-based natural language processing (NLP) system was developed using 480 reports and the information in the corresponding device data was aggregated for the task. We compared the performance of the NLP system with information aggregated from structured device data using the remaining 207 reports.

**Results:** The rule-based NLP system achieved F-measure of 0.92 and 0.98 for ICD therapy and heart rhythm while the performance of aggregating device data was significantly lower with F-measure of 0.78 and 0.45 respectively. Limitations of using only structured device data include no differentiation of real events from management events, data availability, and disparate perspectives of vendor and data granularity while using interrogation reports needs to overcome non-representative keyword/pattern and contextual errors.

**Conclusions:** Extracting phenotyping information from data generated in real-world requires the incorporation of medical knowledge. It is essential to analyze, compare, and harmonize multiple data sources for real-world evidence generation.

## Introduction

Hypertrophic cardiomyopathy (HCM) is the most prevalent inheritable myocardial disease, one of the common causes of sudden cardiac death (SCD) in young people[1-3]. Implantable cardioverter defibrillators (ICDs) are effective in termination of ventricular arrhythmias (ventricular fibrillation or ventricular tachycardia)[4-6], by monitoring heart rhythms and delivering corrective therapies in forms of anti-tachycardia pacing (ATP) and/or shock as needed. For decades, ICDs have been implanted for primary prevention in HCM patients who have high risk of SCD such as family history of SCD[7][8]. However, several studies have reported highly variable data concerning the incidence of appropriate ICD therapy and the complications associated with ICD implantation and discharge[9,10]. Given the lifesaving nature and the undesired risks associated with medical devices, patients with implanted ICDs are closely monitored with clinical consultations occurring whenever ICD therapy is applied.

To assess patient outcomes and to ensure that patients get treatment that is right for them, analyzing device data together with data from electronic health records (EHRs) is necessary[11]. During normal course of operation, these devices generate structured data consisting of numerical data about the ICD therapy itself as well as data about the underlying heart rhythm that triggered therapy. These structured data are stored in the device's own counters and then transmitted to the management system of the healthcare provider. After the occurrence of an interrogated event of ICD device such as an unscheduled visit, in-office visit, emergency phone-call from a patient, or automated telephone alert by the device, an electrophysiologist or trained electrophysiology nurse reviews the electrograms in the time window surrounding the event on top of this structured data on the management system. The reviewer then creates an unstructured ICD interrogation report to interpret ICD therapy, heart rhythm data, past events, along with the device's underlying programming information with the approval of healthcare providers.

Integration or utilization of data from device into EHR has been investigated mainly to access real-word data for healthcare providers and to ensure the safety of patients who have ICD devices through data standardization and improvement of interoperability [11-14].

Given the overall incidence of sudden death is relatively low among HCM patients, it is desired to predict patients with high risk of SCD who can benefit most from ICD implantation. One necessary step to develop predictive models is to accurately assess the adverse cardiac endpoints. In this situation, it would either be patients who died suddenly or patients with ICD devices who had an appropriate discharge for a ventricular arrhythmia. Towards the goal of accurately asserting cardiac endpoints, in this study, we analyzed and compared device data and interrogation reports for extracting phenotyping information of ICD therapy and heart rhythm for a cohort of HCM patients with ICD implanted.

## Materials

**MedTagger: An Information Extraction System**
MedTagger, a system capable of autonomously extracting clinical events from unstructured texts given a clinical dictionary and ruleset[15], has been adapted here to extract electrophysiology experts' interpretation of whether an appropriate ICD therapy was actually delivered as well as the identifiable heart rhythm at the event for comparative purposes with the structured data source.

**Patient Cohort**
For the purposes of this study, the patient cohort used for comparative purposes consisted initially of a cohort of 11,690 patients with a billing diagnosis of HCM (index date: Feb 15, 2019). This HCM cohort was then filtered using structured data from the electronic health record (EHR) of patients that explicitly were billed for implantation and/or occasional treatment or maintenance of an ICD, resulting in a cohort of 1,127 patients being used for this study with institutional review board approval. The specific structured data elements used for filtering purposes is defined in Table 1 below.

| Description | Coding System | Specific codes |
|---|---|---|
| Patients with HCM | ICD 9 (ICD-9-CM) | 425.11, 425.18 |
| | ICD 10 (ICD 10-CM) | I42.2, I42.1 |
| | AND | |
| ICD Diagnosis Codes | ICD 9 (ICD-9-CM) | 996.04, V45.02, V53.32 |
| | ICD 10 (ICD 10-CM) | Z45.02, Z95.810 |

|  |  | OR |
|---|---|---|
| ICD Procedural Codes | Current Procedural Terminology (CPT) | 0319T, 0321T, 0326T, 33223, 33230, 33231, 33240, 33241, 33243, 33244, 33249, 33262, 33263, 33264, 33271, 33272, C1721, C1722, C1882 |
|  | Healthcare Common Procedure Coding System (HCPCS) | C1777, C1895, C1896, C1899 |

Table 1 - Cohort Criteria

## Methods

**Extraction of ICD Therapy and Heart Rhythm**

*Datasets*
For the purposes of developing the MedTagger dictionary used for ICD therapy extraction, we subdivided 687 randomly sampled ICD interrogation reports from the 1,127 patient cohort into a training and test set consisting of 70% being used for training (n=480 cases), and 30% for test (n=207 cases).

*Gold Standard for Training and Test*
The interpretations by electrophysiology experts in the interrogation reports within the dataset were annotated by three independent clinical abstractors. We defined the phenotypes as following:

- ICD Therapy: When the device detects tachyarrhythmia, it delivers the programmed pacing stimulation (ATP therapy) and/or electrical shock (shock) to terminate arrhythmia. In this study, we consider an appropriate ICD therapy is if an interrogation report contains the effective ATP therapy and/or shock for tachyarrhythmia such as "Patient has had 2 VT episodes treated successfully with ATP X 1 each". In case of the failed ICD therapy or ICD therapy for irrelevant arrhythmia, we count as inappropriate ICD therapy. For example, "Patient received 3 failed ATP therapies, and EGMs reviewed and show AF." was inappropriate ICD therapies because of "failed therapies" and "atrial fibrillation". Our annotation task for ICD therapy is to determine whether appropriate ICD therapy is delivered or not.
- Heart Rhythm: If an interrogation report contains any heart rhythm information (i.e. potential underlying reasons for ICD therapy such as "ventricular fibrillation", "ventricular tachycardia", "atrial fibrillation" and "atrial tachycardia"), then we consider it as valid heart rhythm. For example, if "EGMs show VT with rates 150 bpm that accelerated up to 220 bpm." was in an interrogation report; we defined valid heart rhythm identification due to the pattern "EGMs show VT".

The procedure of annotation is as follows: two abstractors independently annotated each interrogation report to identify (1) appropriate ICD therapy and (2) identifiable heart rhythm. Any disagreement of annotations was adjudicated by the third abstractor, a board-certified cardiologist. The inter-annotator reliability of the annotation was calculated with percentage agreement and the Kappa statistic based on a randomly selected sample of 69 interrogation reports (10% of the total 687 interrogation reports).

*System Development*
From the gold standard annotations of the training set, patterns to extract ICD therapy as well as heart rhythm were identified and incorporated into the MedTagger ruleset. The resulting dictionary and the categorization of each pattern for a concept generated from this study are available for public use at https://github.com/RieaM/ICDNLP.git (website). The system extracts fine granularity concepts and normalizes them to the target phenotypes in our evaluation. For example, "ATP therapy" and "shock" concepts were normalized into "ICD therapy". Concepts related to heart rhythms, "ventricular fibrillation", "ventricular tachycardia", "atrial fibrillation" and "atrial tachycardia", were all normalized into "heart rhythm".

*System Evaluation*

The performance of the NLP pipeline and the performance based on information aggregated from structured device data were evaluated against the gold standard and the standard metrics used to evaluate information extraction pipelines, sensitivity, specificity, PPV, NPV, and F1-score were calculated and reported.

# Results

For the inter-annotator reliability of our gold standard, the percent agreement was on 91% and the Kappa statistic was on 0.80 when extracting ICD therapy whereas 91% and 0.82 respectively when extracting the heart rhythm. The gold standard contained 290 appropriated ICD therapy cases (42% of 687) and 327 heart rhythm cases (48% of 687).

We obtained a reasonably high performance for the NLP pipeline to extract the interpreter's assessment on whether ICD therapy was delivered with the corresponding heart rhythm information extracted. A comparison of the data pulled directly from the structured device data fields against their equivalents from the unstructured data from interrogation reports revealed the results as shown in Table 2.

| Data type | Extraction Task | Test Set Performance | | | | |
|---|---|---|---|---|---|---|
| | | Sensitivity | Specificity | PPV | NPV | F-score |
| Structured data | ICD therapy | 0.76 | 0.88 | 0.80 | 0.85 | 0.78 |
| | Heart rhythm | 0.33 | 0.87 | 0.70 | 0.58 | 0.45 |
| Unstructured data | ICD therapy | 0.99 | 0.90 | 0.86 | 0.99 | 0.92 |
| | Heart rhythm | 0.98 | 0.98 | 0.98 | 0.98 | 0.98 |

**Table 2 – Pipeline Performance using Structured or Unstructured Data for ICD Therapy and Heart Rhythm Extraction**

In Table 3, we categorized the major discrepancies (n=14 and n=4 for ICD therapy identification and Heart rhythm identification respectively in test set) the pipeline using unstructured data and the given gold standard.

| Error type | Case | Examples |
|---|---|---|
| Inconsistent format | 7 | 14J: Good detection, no dropped beats. (False Positive of ICD Therapy) |
| Concept validity | 3 | The available EGM shows ventricular sensed beats at 52 bpm with a sudden rate increase to 120 - 136 bpm and a distinct morphology change which could indicate slow NSVT. (False Positive of Heart Rhythm) |
| Non-representative pattern | 5 | Pt broke rhythm before shock delivered. (False Positive of ICD Therapy) |
| Non-representative keyword (e.g. typo) | 2 | Used 200 j externqal (typo of external) shock to convert rhythm to sinus. (False Negative of ICD Therapy) |
| Negation | 1 | The transmission was reviewed, and showed no new VT/VF episodes. (False Positive of Heart Rhythm) |

**Table 3 – Error Analysis**

## Discussion, Challenges, and Future Direction

In this study, we examined heterogeneous data available in the real-world healthcare system, structured data generated from ICD devices and unstructured data from ICD interrogation reports in EHR, to extract phenotyping information regarding ICD therapy and heart rhythm for HCM patients. Our system achieved 0.92 and 0.98 F-score for ICD therapy and heart rhythm identification respectively.

Despite the reliable performance of our system, there are insurmountable challenges and error patterns associated with real-world data analysis of ICD device. The reasons behind the discrepancy between the system and the gold standard for this particular use case are many, and below we have presented several cases and how they relate to the three salient barriers presented in this study.

**Challenges of using structured data**
For the system using structured data only, we observed that there was a significant deterioration in performance both in terms of sensitivity and specificity compared to the system using unstructured data. These results suggest that these device-specific values are not necessarily reflective of the true underlying real-world values, at least within the clinical context pertaining to this particular dataset.

A significant contributor to the lower specificity of device data lies in the fact that it could not distinguish between actual ICD therapy and adjustments of ICD devices due to medical needs. While standardizing structured information (record frequency in device counter in our case), a loss of granularity to adversely affect clinical information (indication of actual ICD therapy or adjustment of ICD device in our case) that may be useful[16-18]. For example, defibrillation threshold (DFT) testing which determines the threshold needed to terminate ventricular fibrillation at the time of device (re-)implantation is commonly practiced. Device functions are assessed to confirm appropriate settings during in-clinic visits on a regular basis. Intervention by physicians via the device (e.g. cardioversion) also frequently happened in emergency situations. Structured data generated through these device adjustments is recorded on the device counter in the same way as structured data generated by real events. (n=9)

Another significant contributor to the lower specificity lies in the fact that the data is tied to the device itself, unavailable or missing information at the point of care: when the device was not accessible, switched, or reset, the historical data was cleared and no longer accessible, thus leading to missed events in the structured data. In particular, the majority of heart rhythms were adjudicated based on reviewing other available material (e.g. historical EGM, transmission via device, downloaded evaluation on the management system of healthcare provider, etc.) and recorded as narrative in ICD interrogation reports. (n=65)

The information extraction and generation strategies implemented out-of-the-box from device represent only a single semantic perspective of the information and are defined by the vendor. This may not always align semantically with the most current interpretations of the underlying unstructured data as used by healthcare providers [18,19]. In this study, the low sensitivity lies in this fact that vendors hold differing definitions of how to define concepts (e.g. therapy or aborted therapy). While structured data contains valid information (frequency in device counter), the reviewer adjudicated as there was no valid therapy based on reviewing other material such as available electrogram (n=13 in ICD therapy identification, n=7 in heart rhythm identification). It is important to note that our emphasis that structured device-generated data represents a single semantic perspective on the underlying data does not mean that it is an inaccurate perspective; indeed the structured data provided by the ICD devices in our case study here can be considered no more or no less valid than those provided by each of our expert interpreters depends on the medical needs. If the device discharged due to detecting an abnormal event but the heart rhythm resolved itself, therapy is diverted (e.g. "Pt broke rhythm before shock delivered."). This, however, is recorded as a positive shock by the vendor, a semantic interpretation that did not agree with that of our human interpreters. The fact that such a difference of concept exists vindicates our assertion that multiple semantic perspectives do exist.

An additional interesting finding is that the scope of information, i.e. the level of data granularity, defined by the device may not correspond to the needs of healthcare providers and depends on the circumstances [20]. Depending on the device, the fine level concepts of "ventricular fibrillation" and "ventricular tachycardia" could be combined as "ventricular rhythm" and "atrial fibrillation" and "atrial tachycardia" potentially represented as "atrial rhythm". Although these fine level concepts may not be essential for the device to function efficiently, the fine granularity of heart rhythm can be useful for long-term medical needs.

**Challenges of using unstructured data**
With all that being said, we also cannot conclusively state that using unstructured data only is the correct approach to take either: an inherent difficulty in computational extraction of information from unstructured data is that it is not 100% accurate, with wide variances in terms of performance. Our system has common NLP challenges such as dealing with non-representative keyword/pattern (e.g. typo), the contextual errors in long sentences (e.g. negation or certainty), etc. In particular, our system had difficulties in recognizing the paragraph that contained device test information because of fragmented sentences or inconsistent formats in ICD interrogation reports (n=8 in ICD therapy identification). For example, the sentence "device charged for second shock" contained appropriate ICD therapy information from the sentence level perspective. However, the adjacent sentence contained the phrase "cardioversion via device" which provided a clue that this therapy was not a real event.

**Challenges of incorporating available heterogeneous datasets with the clinical needs**
In order to harness the heterogeneous dataset for medical purposes, we found it necessary to design with medical knowledge. If we simply combined structured (i.e. information aggregated) and unstructured data (i.e. information extracted using NLP) through Boolean logic OR, the overall performance dropped except for a slight improvement of sensitivity (Table 4).

| Extraction Task | Test Set Performance | | | | |
| --- | --- | --- | --- | --- | --- |
| | Sensitivity | Specificity | PPV | NPV | F-score |
| ICD therapy | 1.00 | 0.79 | 0.75 | 1.00 | 0.86 |
| Heart rhythm | 0.99 | 0.85 | 0.86 | 0.99 | 0.92 |

**Table 4 – Pipeline Performance using Structured and Unstructured Data without medical knowledge**

Deficiency in specificity is due to the format of the structured data, i.e., a data lake that makes no distinction between actual ICD therapy and adjustments of ICD devices. Improvement of sensitivity is evident from our analysis of information extracted. However, relying only on unstructured data will likely not be sufficient. For example, a report with sentences such as "patient was seen for recent arrhythmia with ICD shock" followed by "however, there was no therapy after investigation of device" contradicts the validity of the ICD therapy. Additionally, we found there are 2 cases with structured data providing therapy and heart rhyme information but such information failed to be captured in the interrogation reports.

**Limitations and Future Direction**
This study utilized the ICD interrogation reports to generate as the gold standard for reporting performance that may favorable to unstructured data. The information content supplied by the ICD device differs by vendors. But nevertheless, the conclusions drawn here remain generalizable as the key underlying reasons why such a deficiency exists will remain valid regardless of how that structured data is sourced. This variation in semantic perspectives of interest between diverse dataset from different resources is similar to the variation that exists in the real world. In order to make any clinical decision at the point of care, these semantic perspectives, structured and unstructured, must both be considered when developing and training any system underlying semantics and reasoning of medical needs. Regardless information extraction approaches relying purely on structured data will consistently face significant challenges; there is the value of structured data holds the consistency of method. Implication of our findings is that device data should always be paired with unstructured data for extracting phenotyping information.

# Conclusions

We have analyzed device data from ICD together with EHR data for HCM patients with the ultimate goal of identifying patients who would benefit most from ICD therapy. Through the comparison of structured device data and unstructured interrogation reports for extracting phenotyping information of ICD therapy and heart rhythm, we demonstrated that relying on device data only does not reflect the underlying phenotyping information in the real-world clinical context. Meanwhile, the interrogation reports contain rich and complex information for a diverse range of purposes with different semi-structured formats, causing significant challenges in extracting phenotyping information.

**Acknowledgements:** We would like to thank Luke A. Carlson for insightful comments. The authors also thank Donna Ihrke, Corina Moreno, and Adelaide Arruda-Olson, MD, PhD for manual annotations.
**Contributors:** SM designed the methods and experiments, analyzed the data, and interpreted the results. CS generated dataset. AW implemented the system. SM carried out the experiments. SM drafted the manuscript. SM, HL and RN conceived of the study, and helped to draft the manuscript. All authors read and approved the manuscript.
**Funding**: This work was supported by the Robert D. and Patricia E. Kern Center for the Science of Health Care Delivery Augmented Human Intelligence Care of Tomorrow Award.
**Competing interests:** None.
**Ethics approval:** Mayo Clinic Institutional Review Board.
**Provenance and peer review:** Not commissioned; externally peer reviewed.
**Data sharing statement:** This work has not been published elsewhere and represents original research. Data resources deriving from this article are available at https://github.com/RieaM/ICDNLP.git.